# Octahedral conversion of *a*-SiO$_2$-host matrix by pulsed ion implantation

D. A. Zatsepin[1,2,3], A. F. Zatsepin[2], D. W. Boukhvalov[4,5], E. Z. Kurmaev[1,2], N. V. Gavrilov[6], N.A. Skorikov[1], A. von Czarnowski[7], H.-J. Fitting[7]

[1] M.N. Mikheev Institute of Metal Physics of Ural Branch of Russian Academy of Sciences, 620990 Yekaterinburg, Russia
[2] Institute of Physics and Technology, Ural Federal University, Mira Street 19, 620002 Yekaterinburg, Russia
[3] Institute of Physics, Polish Academy of Sciences, 02-668 Warszawa, Poland
[4] Department of Chemistry, Hanyang University, 17 Haengdang-dong, Seongdong-gu, Seoul 133-791, Korea
[5] Theoretical Physics and Applied Mathematics Department, Ural Federal University, Mira Street 19, 620002 Yekaterinburg, Russia
[6] Institute of Electrophysics, Russian Academy of Sciences, Ural Branch, 62990 Yekaterinburg, Russia
[7] Institute of Physics, University of Rostock, 18051 Rostock, Germany

*This is the abstract. The results of measurements of X-ray photoelectron spectra (XPS) of a-SiO$_2$-host material after pulsed implantation with [Mn$^+$] and [Co$^+$, Mn$^+$]-ions as well as DFT-calculations are presented. The low-energy shift is found in XPS Si 2p and O 1s core-levels of single [Mn$^+$] and dual [Co$^+$, Mn$^+$] pulsed ion-implanted a-SiO$_2$ (E = 30 keV, D = 2·10$^{17}$ cm$^{-2}$) with respect to those of untreated a-SiO$_2$. The similar changes are found in XPS Si 2p and O 1s of stishovite compared to those of quartz. This means that the pulsed ion-implantation induces the local high pressure effect which leads to an appearance of SiO$_6$-structural units in α-SiO$_2$ host, forming "stishovite-like" local atomic structure. This process can be described within electronic bonding transition from the four-fold "quartz-like" to six-fold "stishovite-like" high-pressure phase in SiO$_2$ host-matrix. It is found that such octahedral conversion depends on the fluence and starts with doses higher than D = 3×10$^{16}$ cm$^{-2}$.*

## 1. Introduction

Silicon dioxide is one of the widely spread materials and its different crystalline polymorphs display the useful variety in physical properties (refractive index, density, surface stress, temperature and chemical durability, etc.). It is known, that these polymorphs have an identical chemical composition but dissimilar type of atomic arrangement. The geometry of such kind of atomic arrangement is not only reflected in the symmetry group of a particular polymorph, but also in the isotropy or anisotropy of its physical properties, presence of different types of intrinsic defects, and etc., making these material quite attractive for practical application.

This is well known that in the most of silica polymorphs the Si atom afford tetrahedral coordination with oxygen whereas in some phases (i.e. stishovite) does not. Moreover, the latter is not well described by Mott rule[1], which is pretty good, applied to the "normal" tetrahedral silica modifications. As it follows from the previous studies (i.e. see Ref.[2] and the references therein), the most dissimilarities in physical properties for Si–O polymorphs are explained on the ground of dissimilar Si $3d3s$ and O $2p$ partial DOS contribution to the VB formation in each particular case, but the role of Si $3d$ orbitals in the charge-transfer process along dissimilar Si–O bonding was remained unclear because of the structure complexity of these polymorphs[3]. Nevertheless, it was pointed out that the type of atomic arrangement in unusual allotropic modifications of silicon dioxide may be assumed to be the key for the managed modification of the needed physical property in the $SiO_2$ polymorph that is used as a host-material for ion doping.

One of the possibilities to re-arrange the type of atomic ordering of the host-material is ion implantation. In particularly, it is known that ion implantation into KU-glass host-matrix produces a very high refractive index and increases the density of the material in comparance of other Si–O polymorphs as well as the melting temperature[4]. On the other hand, the type of initial structure transformation of implanted host-matrix, also can be varied significantly, that may lead to some side effects, such as appearance of irregular Si–Si homobonds which are called oxygen deficient centers (ODC), electron $E'$-centers, non-bridging hole-centers, etc. In terms of that, the majority of questions about the influence of particular external treatment on the host-matrix structure and properties still need further study.

In the current paper we present the results of X-ray photoelectron spectroscopy (XPS) in context with the characterization of α-SiO$_2$ host material after pulsed co-implantation with single [Mn$^+$] and double [Co$^+$,Mn$^+$]-ions. Well known, that Co is preferably able to form buried metallic layers in SiO$_2$ (Co-metallization), whereas Mn, being more chemically active, is intensively interacting with oxygen sublattice of the host and limiting oxygen transport caused by implantation (Mn-passivation). Thus it is believed that dual implantation can induce both processes – metallization and passivation – simultaneously. The main idea of our study is to report about the influence of pulsed ion implantation on the local symmetry of constituent atoms in $a$-SiO$_2$ host-matrix without and with thermal annealing. The XPS data are combined for analysis with soft X-ray emission spectroscopy (XES) measurements of ion-implanted $a$-SiO$_2$ [5] and specially performed density functional theory (DFT) calculations of Si $L_{2,3}$ XES and O 1$s$ X-ray absorption spectroscopy (XAS) with accounting matrix elements of transition probability. The performed full electronic structure analysis allows to fix the local octahedral conversion of Si-O units in $a$-SiO$_2$ after single and double ion-implantation.

**2. Experimental and calculation details**

The glassy $a$-SiO$_2$ samples (KU-type silica glass, Type III [6]) before pulsed implantation treatment were visually transparent plane-parallel plates with a surface of an optical quality, measuring (1 × 1) cm$^2$ and a thickness nearly 1 mm. KU glass is a high purity optical silica glass of type III, obtained by hydrolysis technology from silicon tetrachloride vapour in oxygen-hydrogen flame. It has a high homogeneity, very low concentrations of metal impurities and high content of hydroxyl groups (so called "wet" silica glass). These features provide a high radiation-optical stability and a high transparency in the UV and visible regions.

Our samples were implanted by [$Mn^+$] (single ion-implantation) and [$Mn^+$, $Co^+$]-ions (simultaneous dual-ion implantation), using a metal-vapour vacuum arc ion source, developed at the Institute of Electrophysics – Ural Branch of Russian Academy of Sciences (Yekaterinburg, Russia). An applied powdered-arc cathode was made from the sintered manganese powder and sintered mixture of 50 % cobalt and 50 % manganese powders by their weight. The operating pressure in the implantation chamber was not worse than $2.0 \times 10^{-2}$ Pa. The ion energy was set to 30 keV, the pulse duration was 0.4 ms with a current density of 0.6 mA/cm$^2$ and a [Mn, Co]-ion fluency of $2 \times 10^{17}$ cm$^{-2}$. A defocusing of the ion beam was made in order to achieve a lateral uniformity of implanted ions within the host-material. The sample surface temperature during single [$Mn^+$] and simultaneous dual [$Mn^+$,$Co^+$]-irradiation was not exceeding 350 K due to appropriate sample temperature control. No thermal annealing was performed after implantation. A high-purity amorphous *a*-SiO$_2$ was used as a reference XPS spectroscopy standard. It was obtained by an improved melting procedure in order to determine precisely the influence of single and dual-implantation on the *α*-SiO$_2$ host-matrix structure by means of comparison.

X-ray photoelectron spectroscopy (XPS) measurements were made using a PHI XPS Versaprobe 500 spectrometer (ULVAC-Physical Electronics, USA) based on a classic X-ray optic scheme with a quartz monochromator and energy analyzer working in the range of binding energies from 0 to 1500 eV. This apparatus uses electrostatic focusing and magnetic screening to achieve an energy resolution of $\Delta E \leq 0.5$ eV for Al *Kα* radiation (1486.6 eV). The samples were introduced to a vacuum ($10^{-7}$ Pa) for 24 h prior to measurement, and only samples whose surfaces were free from micro impurities were measured and reported herein (surface chemical state mapping attestation). The XPS spectra were recorded using monochromatized Al $K\alpha$ X-ray emission; the spot size was 100 μm dia, and the X-ray power load delivered to the sample was not more than 25 W. Typical XPS signal-to-noise ratios were at least not worth than 11000 : 4. Finally, the spectra were processed using ULVAC-PHI MultiPak Software 9.3 and the residual background (BG) was removed using the Tougard approach with Doniach-Sunjic line-shape asymmetric admixture (well described elsewhere). Well known, that most of the provided background models are self-consistent and they are using Doniach-Sunjic type of line-shapes that are acceptable in most common XPS cases. The advantage of retaining asymmetry in XPS data processing usually strongly apparent when a Tougard BG is used in order to remove the extrinsic contribution to XPS-spectrum of a metal-like or metal-doped materials. Thus, in our case, it is a theoretically based choice. After BG-subtraction, the XPS spectra were calibrated using reference energy of 285.0 eV for the carbon 1$s$ core-level. Exactly such a sequence allows performing much better calibration due to previously removed outer contributions to the XPS line-shape.

Supplementing the results of experimental measurements, we have performed first principle electronic structure calculations based on density function theory (DFT). The calculations of the spectra were performed using the full potential (FP) augmented plane waves + local orbitals (APW + lo) method[7], as implemented in the WIEN2k package[8] in order to solve the scalar relativistic Kohn-Sham equations. We have carried out our calculations using experimentally obtained lattice parameters[9]. An exchange-correlation functional was taken in the function-form type of the Perdew-Burke-Ernzerhof Generalized Gradient Approximation (GGA)[10]. For the calculations of the energetics of Mn incorporation in quartz and stishovite matrixes the pseuadopotential SIESTA code[11] was used with employment of the same GGA functional. A full optimization of the atomic positions was done, during which the electronic ground state was consistently found using norm-conserving pseudopotentials for the cores and a double-$\xi$ plus polarization basis of localized orbitals for Si, Mn and O. The Hellman-Feynman forces and total energies were optimized with an accuracy of 0.04 eV/Å and 1.0 meV, respectively. All calculations were performed in spin polarized mode. For the calculation of both phases (quartz and stishovite), the 72 atoms ($Si_{24}O_{48}$) supercell was used. The calculations of $E_{form}$ formation energies were performed by standard formula: $E_{form} = [E_{def} - (E_{SiO2} - nE_{Si} + nE_{Mn})]/n$, where $E_{def}$ - is the total energy of $SiO_2$ supercell with *n* manganese impurities, $E_{SiO2}$ - is the total energy of supercell of quartz or stishovite and $E_{Si}$, $E_{Mn}$ - are the total energies per atom of silicon and manganese in alpha phase crystals.

### 3. Results and discussions

Figure 1 displays the X-ray photoelectron survey spectrum (XPS survey) of the single and dual-implanted KU-$SiO_2$ sample. Usually, this type of XPS spectra is used in order to perform simple element analysis of the sample under study in order to prove the presence or absence of contaminations. The Survey was measured at least 5 times from different areas of the sample surface for each sample under normal Al *K*α X-ray incident angle with a summation and averaging (so-called Multipoint Mode of Measurements) by means of standard procedures, offered by ULVAC-PHI Software that is processing the XPS data on XPS Versaprobe 500.

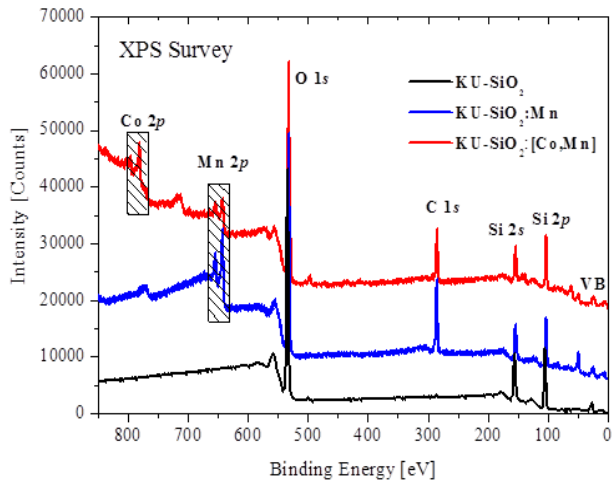

**Figure 1** Comparance of ion-implanted and reference untreated a-SiO$_2$ XPS Survey spectra.

According to the XPS NIST Standard Reference Database[12], there are no extra impurities in the chemical composition of the α-SiO$_2$, SiO$_2$:Mn and SiO$_2$: [Mn,Co] samples except those that are present in the chemical formula. The C 1$s$ = 285 eV signal is arising from ~ 2 nm nano-film of neutral carbon, specially deposited on the sample, in order to perform XPS spectra calibration and to help the neutralization of the surface charging which is common for the wide-gap insulators due to the loss of photoelectrons. So XPS Survey Simple Element Analysis performed allows concluding that the sample is clean and free form unexpected contaminators (see Table 1).

**Table 1** Surface composition of the samples under study

|  | Si, at.% | O, at.% | C, at.% | Mn, at.% | Co, at.% |
|---|---|---|---|---|---|
| a-SiO$_2$ | 39.5 | 58.7 | 1.8 | - | - |
| SiO$_2$ :Mn | 40.5 | 40.3 | 14.4 | 4.8 | - |
| SiO$_2$:Mn+Co | 36.8 | 43.8 | 16 | 1.6 | 1.8 |

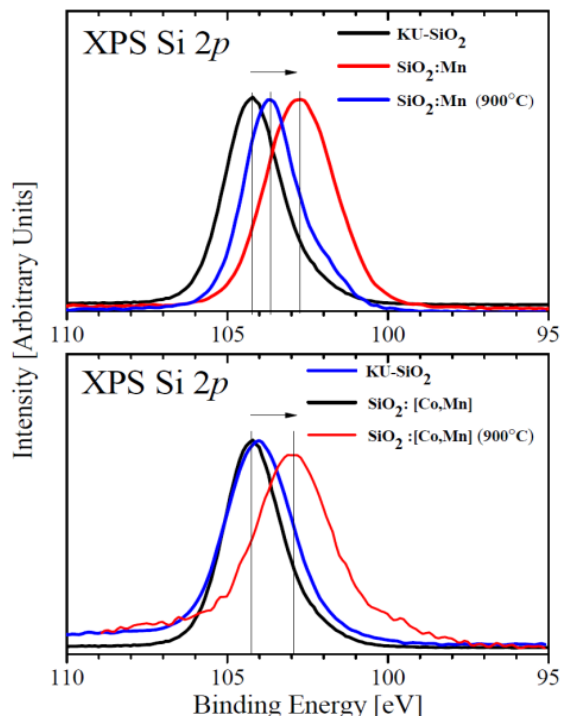

**Figure 2** The high-energy resolved XPS Si 2p spectra for SiO$_2$:Mn (the upper panel) and SiO$_2$:[Mn,Co] samples (the lower panel) without and after thermal annealing (1 hour, 900 $^0$C).

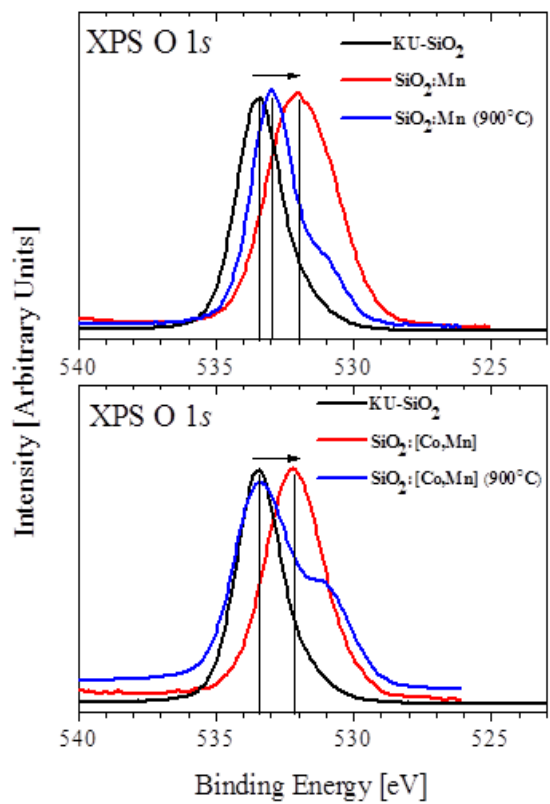

**Figure 3** The high-energy resolved XPS O 1s spectra for SiO$_2$:Mn (the upper panel) and SiO$_2$:[Mn,Co] samples (the lower panel) without and after thermal annealing (1 hour, 900 $^0$C).

The high-energy resolved XPS Si 2$p$ (Fig. 2) and O 1$s$ (Fig. 3) core-level spectra clearly display the low-energy shifts for ion-implanted samples in the same manner as those that were found earlier[13] while comparing natural stishovite and quartz. This means that in our case, the pulsed ion-implantation may induce the local high-pressure on the treated sample surface which leads to the transformation of the initial structural $SiO_4$-units into $SiO_6$ ones in $a$-$SiO_2$ host, probably forming nearly stishovite-like structure. This conclusion is convincingly confirmed by our XPS measurements of the samples with the same implanted composition but with thermal treatment (Figs. 2-3).

Here one can see that the appropriate core-level spectra are exhibiting the reverse shift of XPS Si 2$p$ and O 1$s$ Binding Energy values from implanted and annealed samples to those of untreated reference $\alpha$-$SiO_2$. At the same time, however, seems that this process cannot be recognized by means of XPS as a fully reversible, because even after 1 hour of annealing at 900 $^0$C, for the ion-implanted and annealed samples an XPS sub-band at nearly 532 eV is still present in O 1$s$ core-level (it is quite close in BE-positions with that for "as is" implanted sample), while the main XPS maximum is identical to the reference $a$-$SiO_2$ host (Fig.3, compare spectral parameters for blue, red and black spectra). In other words, this means that some stishovite-like sixfold areas may be still remain in implanted and thermally treated samples. An alternative interpretation for the presence of 532 eV XPS sub-band in the O 1$s$ core-level spectrum of $SiO_2$:[Co,Mn] after annealing might be the possible influence of appeared the fivefold coordinated Si – O defects[14] due to uncompleted reverse transition – from sixfold or "stishovite-like" to fourfold or "quartz-like" structure. Such an interpretation also seems to be plausible, because as it is well known, that spectrally the presence of defects in the structure of a material might be detected by broadened peaks due to their irregular geometry and hybridization between neighboring atoms. And in our case the FWHM of 532 eV XPS sub-band in O 1*s* core-level spectrum of $SiO_2$:[Co,Mn] well agrees with the previous notes. But in order to answer the question about the origin of this sub-band in a more detailed manner the further additional study is needed.

Figure 4 presents the comparison of Si $L_{2,3}$ X-ray emission spectra (Si $3s3d$ □ $2p$ electron transition, for single and dual-ion implanted *a*-$SiO_2$ [5]) with that for quartz and stishovite[15,16]. One can see from the upper panel of Fig. 4, that an implantation causes the noticeable changes in relative intensity of sub-bands which are ranging at relatively low-emission energies (84–93 eV). The spectral parameters of these sub-bands are strongly affected by applied ion fluency becoming clearly noticeable when D = $3\times10^{16}$ cm$^{-2}$ and higher. The similar changes in relative intensities of Si $L_{2,3}$ XES sub-bands were also found in appropriate XES spectra of stishovite and quartz (Fig. 4, the lower panel).

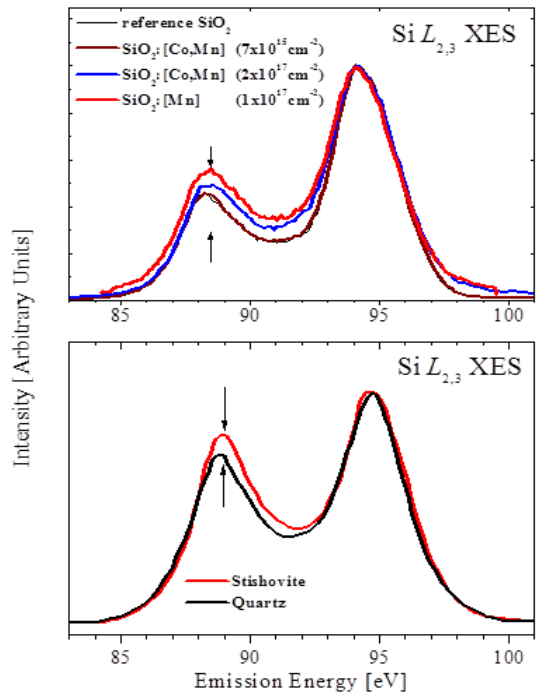

**Figure 4** XES Si $L_{2,3}$ of ion-implanted *a*-$SiO_2$ [5] (the upper panel) compared with those of quartz and stishovite (the lower panel)[12,13].

Figure 5 presents PDOS curves and theoretically simulated Si $L_{2,3}$ spectra for quartz and stishovite after in band structure calculations. Relatively wide valence band of quartz consists of two sub-bands that are separated by an energy gap of approximately 1 eV. This gap separates the O 2p π-nonbonding orbitals and the O 2p–Si 3s, 3p, and 3d σ-bonding orbitals and originated as consequence of twofold O coordination [19]. A possible explanation of calculated energy gap in the band structure calculations and the absence of a gap in the experimental X-ray emission spectra might be based on many-electron effects that manifest themselves in the experiment and are not involved into the band calculations. Si partial densities of states display that the *s*-symmetry Si states are mainly concentrated at the bottom of valence band while the *d*-symmetry Si states are located at top. The structures of valence band for stishovite and quartz are not similar one because of the gap absence between bonding and unbonding oxygen states that is a consequence of threefold coordination of O-atoms by Si.[19] The reduction of O–O distances in stishovite compared to quartz leads to the broadening of O 2p bands and another hybridization with Si *s,p* and *d*-bands. The bottom panel of figure 5 presents theoretical XES Si $L_{2,3}$ spectra for quartz and stishovite obtained in Density Functional Theory calculations where the matrix elements of transition probability were involved. Theoretically simulated spectra reproduce on the whole the experimentally determined differences between Si $L_{2,3}$ spectra of quartz and stishovite (increase of intensity of low-energy maximum and broadening of spectra of stishovite compared to quartz). The comparison of theoretical spectra with PDOS curves shows that the main peak at -3 eV is originated from Si *d*, and second maximum at -9 eV is originated from Si *s*-states.

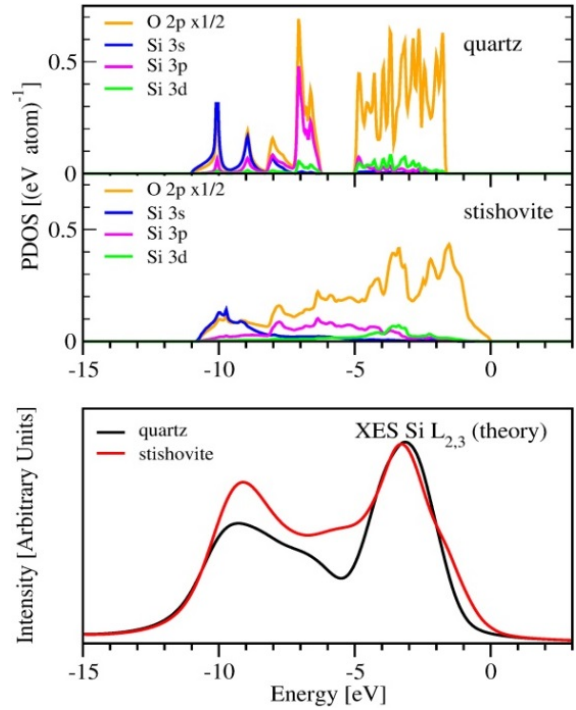

**Figure 5** Partial Density of States (PDOS) and DFT-calculated XES Si $L_{2,3}$ of quartz and stishovite.

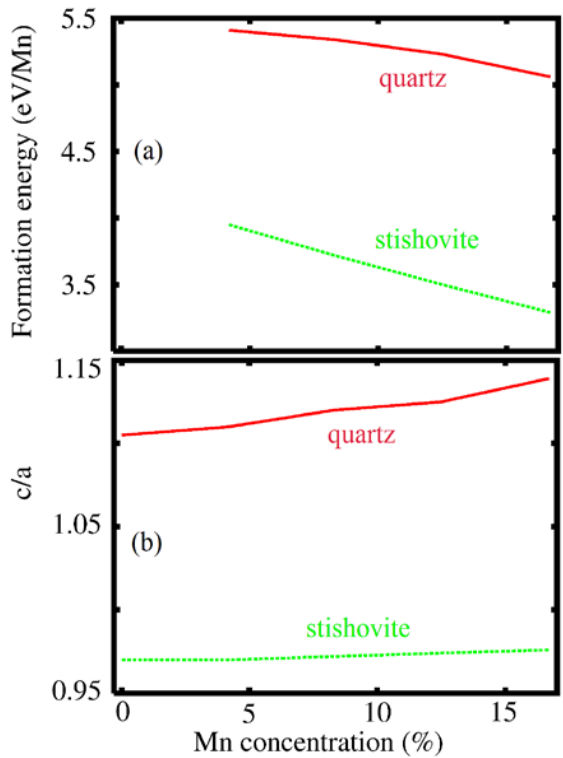

**Figure 6** The formation energies and ratio of *c/a* lattice parameters as function of manganese concentration.

To estimate the possibility of quartz to stishovite transformation as results of Mn incorporation we have calculated the formation energies and changes of ratio of *c/a* crystal lattice parameters as function of Mn concentration. The results of calculations demonstrate that for stishovite phase of $SiO_2$ the formation are about 1.5 eV lower and faster decay with the grown of Mn concentration than in quartz phase takes place (Fig. 6a). Thus one can conclude that transformation of quartz to stishovite in Mn-rich areas is extremely energetically favorable. The calculation of the changes of lattice parameters ratio shows that the presence of Mn impurities provides increasing of *c/a* ratio (Fig. 6b) with the impurities concentration in quartz in contrast of that for stishovite matrix where the presence of Mn impurities provides only negligible changes in *c/a* ratio. So, we can speculate that the presence of Mn impurities destabilize this phase by distortion of crystal lattice. For the check of the effect of magnetic ordering to the energetics of these processes we calculate the total energies for the ferromagnetic and antiferromagnetic configurations of spins on Mn for each studied configurations and finds that antiferromagnetic configurations always have lower total energies in order of 1 meV/Mn. Thus we can conclude absence of any magnetic order in Mn doped $SiO_2$.

Finally, our XPS-XES-XAS studies provide a conclusive evidence for a fourfold, "quartz-like" to a sixfold, "stishovite-like" electronic Si–O bonding transition in implanted KU-$SiO_2$ host-matrix. The result obtained is of a concrete applicative importance because the reported phenomena might be exactly used for altering the most of physical properties of the materials in the set of Si – O polymorphs. The observed ion-implantation induced electronic bonding transition in KU-$SiO_2$ host-matrix was found not to be completely reversible upon permanent value of temperature treatment, thus allowing to suppose that another mode or/and type of annealing have to be employed for that.

## 4. Conclusions

An XPS study after high-intensity flux ($10^{17}$ cm$^{-2}$) of single and dual implantation with Co$^+$ and Mn$^+$ ions into KU quartz glass was made. It was found that the XPS Si 2$p$ and O 1$s$ spectra of ion implanted $a$-SiO$_2$ become very similar with those of natural stishovite. The reason for this change of the electronic structure is the strong re-arrangement of atomic structure caused by the high local-pressure of single and co-implantation. Thus, taking into account all the aforesaid, we assume that a part of the common SiO$_4$ coordination of the amorphous host-matrix transforms into SiO$_6$ "stishovite-like" structural units (high-pressure phase) with the residual SiO$_4$ units.

**Acknowledgements** XPS measurements and DFT calculations are supported by Russian Science Foundation (Grant 14-22-00004). The preparation of $a$-SiO2 samples and ion implantation treatment are supported by Deutsche Forschungsgemeinschaft - DFG (German Research Foundation Project FI 497/15-1) and Russian Ministry of Science and Education (Government Task).

## References

[1] V.A. Gritsenko, Phys.-Usp. **52**, 869, 2009.

[2] E. Gnani, S. Reggiani, R. Colle, M. Rudan, IEEE Trans. Electron Dev. **47**, 1795, 2000.

[3] S. S. Nekrashevich, V.A. Gritsenko, Physics of the Solid State **56**, 209, 2014.

[4] P. D. Townsend, P.J. Chandler, L. Zhang, Optical effects of Ion Implantation, England:Cambridge Press, 296, 2006.

[5] R.J. Green, D.A. Zatsepin, D.J.St. Onge, E.Z. Kurmaev, N.V. Gavrilov, A.F. Zatsepin, A. Moewes, J. Appl. Phys. **115**, 103708, 2014.

[6] R. Brückner, Encyclopedia of Appl. Phys. **18**, S101,1997.

[7] G.K.H. Madsen, P. Blaha, K. Schwarz, E. Sjöstedt, L. Nordström, Phys. Rev. B **64**, 195134, 2001.

[8] P. Blaha, K. Schwarz, G. Madsen, D. Kvasnicka, J. Lutiz, Computer code wien2k, an augmented plane wave + local orbital program for calculating crystal properties, (Karlheinz Schwarz, Technische Universität Wien, Austria, 1999.

[9] T. Yamanaka, R. Kurashima, J. Mimaki, Zeitschrift fur Kristallographie, **215**, 381, 2000.

[10] J. P. Perdew, K. Burke, M. Ernzerhof, Phys. Rev. Lett. **14**, 3865, 1996.

[11] J.M. Soler, E. Artacho, J. D. Gale, A. García, J. Junquera, P. Ordejón, D. Sánchez-Portal, J. Phys.: Condens. Matter, **14**, 2745, 2002.

[12] NIST Standard Reference Database, version 4.1, http://srdata.nist.gov/xps/ (called 2014-05-28).

[13] I. Sunagawa, H. Iwasaki, F. Iwasaki, Growth and Morphology of Quartz Crystals: Natural and Synthetic, Terrapub, Tokyo, ISBN 978-4-88704-146-2, 2009.

[14] J.F. Stebbins, S. Kroeker, S.K. Lee, T.J. Kiczenski, J. Non-Cryst. Solids **275**, 1, 2000.

[15] G. Wiech, Solid State Communications **52**, 807, 1984.

[16] A. Šimůnek, J. Vackář, G. Wiech, J. Phys.: Condens. Matter **5**, 867, 1993.

[17] J.-F. Lin, H. Fukui, D. Prendergast, T. Okuchi, Y.Q. Cai, N. Hiraoka, C.-S. Yoo, A. Trave, P. Eng, M.Y. Hu, P. Chow, Phys. Rev. B **75**, 012201, 2007.

[18] D.A. Zatsepin, V.R. Galakhov, B.A. Gizhevskii, E.Z. Kurmaev, V.V. Fedorenko, A.A. Samokhvalov, S.V. Naumov, R. Berger, Phys. Rev. B **59**, 211, 1999.

[19] K. Seino, F. Bechstedt, P. Kroll, Nanotechnology **20**, 135702, 2009.